\documentclass[twocolumn,showpacs,preprintnumbers,amsmath,amssymb]{revtex4}
\usepackage{graphicx}% Include figure files
\usepackage{dcolumn}% Align table columns on decimal point
\usepackage{bm}% bold math

\begin{document}

\title{Quantization of static inhomogeneous spacetime}% Force line breaks with \\

\author{Shintaro Sawayama}
 \affiliation{Department of Physics, Tokyo Institute of Technology, Oh-Okayama 1-12-1, Meguro-ku, Tokyo 152-8550, Japan}%Lines break automatically or can be forced with \\
\email{shintaro@th.phys.titech.ac.jp}

\date{\today}% It is always \today, today,
\begin{abstract}
In this paper we can solve a Wheeler-DeWitt equation of the some inhomogeneous spacetime models as a local solution.
From the previous study of up-to-down method we derived the static restriction relating the problem of the time.
Although static restriction does not commute with the general Hamiltonian constraint, the Hamiltonian constraint of some mini-superspace models 
commute with static restriction.
We can quantize such inhomogeneous models.
With obtained result we can success to simplify the general local Hamiltonian constraint.
\end{abstract}

\maketitle
\section{Introduction}\label{sec1}
The one of the main issue of the quantum gravity \cite{De} is the quantization of the inhomogeneous spacetime.
There are other problems such as problem of the norms and problem of the time.
The mini-super approach or loop approach \cite{Rov}\cite{As}\cite{Thi} can be quantize only homogeneous spacetime.
Although the loop quantum gravity treats quantization of inhomogeneous spacetime, the Hamiltonian constraint of the inhomogeneous spacetime still not solved.

In the quantization of the spacetime, whether we use a general metrics or static metrics does not simply the Hamiltonian constraint.
However, the study of Wheeler-DeWitt equation i.e. up-to-down method \cite{Sa}\cite{Sa2} become clear that if we treat static mini-superspace there appears 
a additional constraint that we call static restriction.
Although in the previous paper we derived incorrect solution of the Wheeler-DeWitt ,the previous work is relating to the problem of the time \cite{IK}.
So our work is relating to the problem of the time \cite{IK}\cite{SS}.

Usually static restriction and the Hamiltonian constraint does not commute.
However, the Hamiltonian constraint of some inhomogeneous model commutes with static restriction 
because the static metric create a time evolution even if the metric does not depend on time.
We treat the inhomogeneous model of the metric diagonal space and $q_{ii}$ has the support of $x_j,x_k$ for $j,k\not=i$.
Or we can say we treat inhomogeneous model of diagonal metric model and the model that $q_{ii}$ does not depend on $x_i$.
Then we can easily quantize some inhomogeneous mini-superspace models.
In this paper we show two type of solution of the Hamiltonian constraint with static restriction.
We treat these problems in the basis of LQG or metric diagonal models.

Note that we do not use the term LQG as Haag's textbook \cite{Haag} or a loop quantum gravity.
We use this term as the local solution of the Wheeler-DeWitt equation.
And we comment on the term inhomogeneous mini-superspace model.
From now on mini-superspace model is used to only the homogeneous case,
however, we enlarge this term to the inhomogeneous case.
We use this term as quantization of the Gowdy model \cite{GM}.

In this paper we first introduce what is LQG in this paper.
Although there appears boundary condition problem, we ignore it in this paper.
In the next step we introduce static restriction with up-to-down method.
And in the next step we quantize locally inhomogeneous spacetime.

In section \ref{sec2} we explain what is the local quantum gravity in this paper.
In section \ref{sec3} we treat the problem of the time and derive static restriction.
In section \ref{sec4} we solve some static inhomogeneous spacetime and then 
we can simplify the usual local Hamiltonian constraint.
There are two inhomogeneous models which can be quantized by the up-to-down method.
And in section \ref{sec5} we summarize our result and we note further motivation.
\section{Local solution of the quantum gravity}\label{sec2}
What we call local quantum gravity is simple quantization of the spacelike metric diagonal model with boundary condition.
Note that the timelike metric is not diagonal because of diffeomorphism constraints.
We use a fact that the spacelike metric become diagonal at the hyperspace $\Sigma$ by the local coordinate transformation.
The hyperspace can be decomposed such that the spacelike metric become diagonal.
Although the local solution is not complete yet, we only treat a spacelike metric diagonal mini-superspace, the reader can skip this section.
The Hamiltonian constraint of the metric diagonal space is
\begin{eqnarray}
\sum_{ij}\frac{1}{2}\frac{\delta ^2}{\delta \phi_i\delta \phi_j}+\sum_{i\not=j}(\phi_{i,jj}+\phi_{j,i}\phi_{i,i})e^{\phi_i}=0.
\end{eqnarray}
And the diffeomorphism constrain is 
\begin{eqnarray}
\frac{\partial}{\partial x_i}\bigg( \frac{\delta}{\delta q_{ii}}\bigg) .
\end{eqnarray}
Although the off-diagonal components does not enter in this formulation,
the off-diagonal components are contained in local coordinate.
The different point of usual formulation is appearance of the boundary condition.
So in this formulation we should solve Hamiltonian constraint with boundary condition as
\begin{eqnarray}
\sum_{ij}\frac{1}{2}\frac{\delta ^2}{\delta \phi_i\delta \phi_j}+\sum_{i\not=j}(\phi_{i,jj}+\phi_{j,i}\phi_{i,i})e^{\phi_i}+{\rm bound}=0.
\end{eqnarray}
Here, bound means boundary condition term.
However, we ignore this term in this paper for simplicity.

If we solve this Hamiltonian constraint we recover off-diagonal component in the local coordinate.
However, the consistency of this method is further work.
We should check the consistency between two state one is the state that at first time we solve spacetime grovel with off-diagonal metric 
components and 
look the state in local coordinate with only diagonal components of the metric and 
the state at first time we solve locally as in this paper.
However, we can not check this point because the usual Hamiltonian constraint still not solved.

\section{static restriction}\label{sec3}
In this section we explain what is static restriction.
In the subsection \ref{sec3-a} we explain the procedure and in section \ref{sec3-b} we explain the static restriction.
\subsection{Up-to-down method}\label{sec3-a}
In this section we rewrite what is up-to-down method, and some sentence is same to previous paper.

We start by introducing the additional dimension which is an external euclidean time with positive signature, 
and thus create an artificial enlarged functional space corresponding to this external time.
We write such external dimension as $s$.
The action may be written as
\begin{eqnarray}
S=\int _{M\times s}{}^{(5)}RdMds.
\end{eqnarray}
Where ${}^{(5)}R$ is the 5-dimensional Ricci scalar, 
built from the usual 4-dimensional metric and external time components. 
Rewriting the action by a 4+1 slicing of the 5-dimensional spacetime with lapse functionals given by the $s$ direction, 
we obtain the 4+1 Hamiltonian constraint and the diffeomorphism constraints as,
\begin{eqnarray}
\hat{H}_S\equiv \hat{R}-\hat{K}^2+\hat{K}^{ab}\hat{K}_{ab} \\
\hat{H}_V^a\equiv \hat{\nabla} _b(\hat{K}^{ab}-\hat{K}\hat{g}^{ab}),
\end{eqnarray}
where a hat means 4-dimensional, 
e.g. the $\hat{K}_{ab}$ is extrinsic curvature defined by $\hat{\nabla}_a s_b$ and $\hat{K}$ is its trace, 
while $\hat{R}$ is the 4-dimensional Ricci scalar, 
and $\hat{\nabla} _a$ is the 4-dimensional covariant derivative. \\
 \\
{\it Definition. } The artificial enlarged functional space is defined by 
$\hat{H}_S|\Psi^{5} (g)\rangle =\hat{H}_V^a|\Psi^{5} (g)\rangle =0$, 
where $g$ is the 4-dimensional spacetime metrics 
$g_{\mu\nu}$ with ($\mu =0,\cdots ,3$).
We write this functional space as ${\cal H}_5$.
 \\ 
 
Here, the definition of the canonical momentum $P$ is different from the usual one. 
Note in fact that the above state in ${\cal H}_5$ is not the usual 5-dimensional quantum gravity state, because the 4+1 slicing is along the $s$ direction.
This is why we call this Hilbert space as artificial functional space. 
It is not defined by $\partial {\cal L}/(\partial dg/dt)$ but by $\partial {\cal L}/(\partial dg/ds)$, where 
${\cal L}$ is the 5-dimensional Lagrangian. 

In addition, we impose that 4-dimensional quantum gravity must be recovered from the above 5-dimensional action.
The 3+1 Hamiltonian constraint and diffeomorphism constraint are,  
\begin{eqnarray}
H_S\equiv {\cal R}+K^2-K^{ab}K_{ab} \\
H_V^a\equiv D_b(K^{ab}-Kq^{ab}).
\end{eqnarray}
Here $K_{ab}$ is the usual extrinsic curvature defined by $D_at_b$ 
and $K$ is its trace, while ${\cal R}$ is the 3-dimensional Ricci scalar, 
and $D_a$ is the 3-dimensional covariant derivative.
Then we can define a subset of the auxiliary Hilbert space on which the wave functional satisfies the usual 4-dimensional constraints. 
In order to relate the 4 and 5 dimensional spaces we should define projections.
\\ \\
{\it Definition.} The subset of ${\cal H}_5$ in which the five dimensional quantum state satisfies the
extra constraints $H_SP|\Psi ^5(g)\rangle=H_V^aP|\Psi ^5(g)\rangle =0$
is called ${\cal H}_{5lim}$, where $P$ is the projection
defined by
\begin{eqnarray}
P:{\cal H}_5 \to L^2_4 \ \ \ 
\{ P|\Psi^5(g)\rangle=|\Psi^5(g_{0\mu}={\rm const})\rangle \} ,
\end{eqnarray}
where $L^2_4$ is a functional space.
And ${\cal H}_4$ is the usual four dimensional state with the restriction that 
$H_S|\Psi^4(q)\rangle=H_V^a|\Psi ^4(q)\rangle=0$.
Here $q$ stands for the 3-dimensional metric $q_{ij}(i=1,\cdots ,3)$, and
$P^{\dagger}$ is defined by
\begin{eqnarray}
P^{\dagger}:{\cal H}_{5lim} \to {\cal H}_4 .
\end{eqnarray}
\\ 

The enlargement is carried by the multiplying the arbitrary functional $f[g_{\mu\mu},g_{\mu x_i}]$ for usual quantum gravity state of 4-dimensional.
Otherwise the measure of the projection is zero.
This enlargement solves the measurement problem of the projection.

From now on we consider the recovery of the 4-dimensional vacuum quantum gravity from the 5-dimensional wave functional.
We assume that the constraint,
\begin{eqnarray}
\hat{R}|\Psi^{5}(g)\rangle =0
\end{eqnarray}
holes. Here $\hat{R}$ is the operator,
corresponding to the 4-dimensional Ricci scalar as a Dirac operator.
Then the modified Hamiltonian constraint for the 5-dimensional 
quantum state which contains the 4-dimensional Einstein gravity becomes,
\begin{eqnarray}
-m\hat{H}_S:= -\hat{K}^2+\hat{K}^{ab}\hat{K}_{ab}+\dot{q}_{ij}P^{ij},
\end{eqnarray}
where $m\hat{H}$ is called modified Hamiltonian constraint simplified by using 3+1 constraint equations.
There is the theoretical branch in using the Dirac constraint or Hamiltonian and diffeomorphism constraint.
To use the Dirac constraint at this point create the additional constraint which restrict the state to the static.

Finally, the simplified Hamiltonian constraint in terms of the canonical representation becomes
\begin{eqnarray}
m\hat{H}_S =(-g_{ab}g_{cd}+g_{ac}g_{bd})\hat{P}^{ab}\hat{P}^{cd}-\dot{q}_{ij}P^{ij}.
\end{eqnarray}
The magic constant factor $-1$ for the term $g_{ab}g_{cd}$ is a consequence of the choice of dimensions for ${\cal H}_5,{\cal H}_4$.
Here $\hat{P}^{ab}$ is the canonical momentum of the 4-dimensional metric 
$g_{ab}$, that is $\hat{P}^{ab}=\hat{K}^{ab}-g^{ab}\hat{K}$. 
And as we mentioned above, this canonical momentum is defined by the external time and not by the usual time.
We does not write $\dot{q}_{ij}$ by the commutation relation of the Hamiltonian constraint and canonical momentum at this step.

We now give a more detailed definition of the artificial functional space as follows: \\ \\ 
{\it Definition.} The subset ${\cal H}_{5(4)}\subset {\cal H}_5$ is defined by the constraints,  
$ \hat{R}|\Psi ^5(g)\rangle =0$, 
and we write its elements as $|\Psi ^{5(4)}(g)\rangle$.
We also define a projection $P^*$ as
\begin{eqnarray}
P^* : {\cal H}_{5(4)} \to {\cal H}_{4(5)} \nonumber \\
  \{ P^*|\Psi ^{5(4)}(g)\rangle =|\Psi ^{5(4)}(g_{0\mu}={\rm const})\rangle 
=: |\Psi ^{4(5)}(q)\rangle \} ,
\end{eqnarray} 
where ${\cal H}_{4(5)}$ is a subset of ${\cal H}_4$.
\\ 

We notice the projection $P$ and $P^*$ is almost all same. 
However, we use the different symbol because the projected functional space is different.\\ 

{\it Theorem.} In this method, in ${\cal H}_4$ additional constraint $m\hat{H}_SP^*=0$ appears.
\subsection{Static restriction}\label{sec3-b}
The theorem of the end of section II is the main result of static restriction.
If we write this additional constraint in terms of operators, we obtain
\begin{eqnarray}
(q_{ij}q_{kl}-q_{ik}q_{jl})\frac{\delta}{\delta q_{ij}}\frac{\delta}{\delta q_{kl}}=0
\end{eqnarray}
This is the static restriction.
The static restriction is not consistent with the usual Hamiltonian constraint.
However, there is reason why static restriction and Hamiltonian constraint does not commute.
The reason may be trivial because all the space which is quantized is not static state.
If the all the spacetime can be quantized simultaneously, the evolution of the spacetime vanishes and it reduces contradiction.
The static state is very special state of the quantum state.
Although consistency is broken, we treat 4-dimensional gravity not higher dimensional gravity.
We can find the consistent space whose Hamiltonian and static restriction does commute.

The broken of the consistency comes from at the point that we use Dirac constraint.
Whether we use Dirac constraint or Hamiltonian and diffeomorphism constraint is different.
If we treat quantization of static metric space, the corresponding quantum state has time evolution.
This is the different point of classical gravity and quantum gravity.

In the local quantum gravity, the static restriction becomes as,
\begin{eqnarray}
\sum_{i\not= j}\frac{\delta^2}{\delta\phi_i\delta\phi_j}=0.
\end{eqnarray}
At this point we treat metric diagonal mini-superspace.
We use static restriction in the framework of mini-superspace model.
So we can ignore the $P^{ij}$ for $i\not=j$ components at this point. 
We use above restriction as a ansatz.

The broken of the consistency also comes from the point that we treat metric diagonal mini-superspace model.

\section{Quantization of the static inhomogeneous spacetime}\label{sec4}
We treat a mini-superspace such that,
\begin{eqnarray}
q_{ab}=\begin{pmatrix}
q_{11}(x_2,x_3) & 0 & 0 \\
0 & q_{22}(x_1,x_3) & 0 \\
0 & 0 & q_{33}(x_1,x_2)
\end{pmatrix}.
\end{eqnarray}
Then the Hamiltonian constraint corresponding to this mini-superspace is,
\begin{eqnarray}
\sum_i\frac{\delta^2}{\delta \phi_i\delta\phi_j}+8\phi_{i,jj} e^{\phi_i}=0.
\end{eqnarray}
In this case the Hamiltonian and static restriction does commute and we treat usual 4-dimensional quantum gravity.
The corresponding space of the above metric to the classical space may not exist.
However, to solve this Hamiltonian constraint, we can simplify usual Hamiltonian constraint.
And in the next example we can find a classical corresponding space.
 
This Hamiltonian constraint commute with static restriction because the the Hamiltonian constraint does not contain cross terms of $\phi_i$ and $\phi_j$ 
in the non-linear terms and we assume static restriction is satisfied on the state.
However, in this inhomogeneous model, the Hamiltonian and static restriction commute.
We should comment on the fact that the diagonal metric space has usually dynamics or time evolution.
Because if $q_{ii}$ has support on $x_i$, the cross terms appear in the Hamiltonian constraint.
And we can carry on simultaneous quantization.
Usually the state should be a functional of $q_{ij}$ and $q_{ij}$ has the support of the $t,x_1,x_2,x_3$.
In this paper we fix a time constant gauge and we assume $q_{ii}$ does not depend on $x_i$.

At first we comment on the previous work.
In the our previous work, we derived one ordinary differential constraint equation with non-linear term.
That is the Hamiltonian constraint of the mini-superspace of $\phi_{i,i}=0,\phi_j={\rm const},\phi_k={\rm const}$,
\begin{eqnarray}
\frac{\delta^2}{\delta a_i^2}+4\partial^j\partial_j\ln \hat{a_i}=0 \ \ \ {\rm for \ \ some} \ \ i
\end{eqnarray}
Here, $a_i=g_{ii}^{1/2}$ and hat means operator.
The above equation is created from the Eq.(3) with simplification by static restriction
\begin{eqnarray}
\sum_{i\not= j}\frac{\delta^2}{\delta\phi_i\delta\phi_j}=0,
\end{eqnarray}
where, $g_{ii}=e^{\phi_i}$.
The static restriction to the state is ansatz and it commute with Hamiltonian constraint of above mini-superspace Hamiltonian constraint.
We can solve this equation easily if we ignored the operator ordering.
And the solution is,
\begin{eqnarray}
\cos (2\int (\partial^j\partial_j\ln a_i)^{1/2}\delta a_i) \ \ \ {\rm for} \ \ \  i\not= j.
\end{eqnarray}
If we acted it momentum constraint to the above solution, we obtain
\begin{eqnarray}
\sum_{i\not= j}\phi_{i,jj}=c_i(x_j,x_k)^2 \ \ \ {\rm for} \ \ \ j,k\not= i,
\end{eqnarray}
Because we use the static restriction as ansatz of state momentum constraint also commute with static restriction.
Then the state is,
\begin{eqnarray}
|\Psi^4(q)\rangle =\cos (2\int c_i(x_j,x_k)\delta a_i) \ \ \ {\rm for} \ \ \  j,k\not= i.
\end{eqnarray}
The important point is the Eq.(22) not the solution.
The cosine wave comes from approximation.
We use this simple example in later.

We can explain the reason why we insert the operator ordering term as previous paper.
There is point like quantization which starts with the action decomposition as,
\begin{eqnarray}
S=\int RdM\to \int dt \sum_{x^{(i)}} R_i[g^{(i)}_{\mu\mu}(x^{(i)})].
\end{eqnarray}
Here we use assumption of spacetime is separated so the integration become summation.
Then the equation corresponding Eq.(3) becomes as,
\begin{eqnarray}
\frac{\partial^2}{\partial a_i^2}+4\partial^j\partial_j\ln \hat{a_i}=0 \ \ \ {\rm for \ \ some} \ \ i.
\end{eqnarray}
And we can ignore the differential term of $x_j$.
This solution is 
\begin{eqnarray}
\cos (2c_ia_i). 
\end{eqnarray}
Although we ignore the deferential term of $x_i$ such as $\partial^j\partial_j\ln \hat{a_i}$, 
we do not ignore the deferential term of $a_i$.
So the solution is cosine wave.
If we take a point like limit of Eq.(7), we obtain the same result without constant factor.
So we choose a operator ordering term to that the both limit become same.

In the next step we enlarge this mini-superspace for the $\phi_{i,i}=0$ case.
Then the Hamiltonian constraint of the mini-superspace of (20) is simplified by using static restriction and becomes as,
\begin{eqnarray}
\sum_i\frac{\delta^2}{\delta \phi_i^2}+\sum_{i\not= j}\phi_{i,jj}e^{\phi_i}=0 .
\end{eqnarray}
This Hamiltonian constraint commute with static restriction if we assume that the state satisfy static restriction.
Then we can create parameter separated solution as
\begin{eqnarray}
|\Psi^4 (q)\rangle =h[g_{11}]h[g_{22}]h[g_{33}]
\end{eqnarray}
where $h$ is the functional.
This parameter decomposition is also assumption.
We discus the viability of this assumption.
If we act it to the Hamiltonian constraint of mini-superspace (20) we can obtain ordinary differential equation which is Eq.(22).
So this assumption is valid if we only consider the Hamiltonian constraint.
Although the Hamiltonian constraint contains second order derivative, the cross term does not appear because of static restriction assumption.
Because of the parameter separation we fix a gauge.
However, what gauge we fix is further work.
The measure of state may be empty because of this gauge fixing.
Our method to quantize this mini-superspace model has following three steps.
First we solve the Hamiltonian constraint and next time we solve diffeomorphism constraint and finally we use a static restriction.
However, if we change a step to that in the second time we use a static restriction and finally we use diffeomorphism constraint,
the obtained result is same.
Because, we use this parameter separated assumption, the solution is not the general solution but the special solution. 
And the solution of the Hamiltonian constraint is
\begin{eqnarray}
\cos (2\sum_i\int (\partial^j\partial_j\ln a_i)^{1/2}\delta a_i) \ \ \ {\rm for} \ \ \  i\not= j.
\end{eqnarray}
Then diffeomorphism constraint is three set of
\begin{eqnarray}
\sum_{i\not= j}\phi_{i,jj}=c_i(x_j,x_k)^2 \ \ \ {\rm for} \ \ \ j,k\not= i.
\end{eqnarray}
Then the state becomes as,
\begin{eqnarray}
|\Psi^4(q)\rangle =\cos (2\sum_i\int c_i(x_j,x_k)\delta a_i) \ \ \ {\rm for} \ \ \  j,k\not= i.
\end{eqnarray}
Because of the static restriction we can derive,
\begin{eqnarray}
c_1(x_2,x_3)=\frac{-c_2(x_1,x_3)c_3(x_1,x_2)}{c_2(x_1,x_3)+c_3(x_1,x_2)}.
\end{eqnarray}
If the above equation is consistent $c_2$ is only the function of $x_3$ and $c_3$ is only the function of $x_2$.
Then we obtain relation of 
\begin{eqnarray}
c_1(x_2,x_3)=\frac{-c_2(x_3)c_3(x_2)}{c_2(x_3)+c_3(x_2)}.
\end{eqnarray}
With these function of $c_i$, we can rewrite the state functional as
\begin{eqnarray}
\cos (2\int c_1(x_2,x_3)\delta a_1+\int c_2(x_3)\delta a_2+\int c_3(x_2)\delta a_3 ) 
\end{eqnarray}
There are three permutation of in the selection of $c_i$ and there are $2^3$ selection of signature, 
so there are 24 basis.
We can easily enlarge this state with time evolution as,
\begin{eqnarray}
c_1(x_2,x_3)\to c_1(t,x_2,x_3) \nonumber \\
c_2(x_3) \to c_2(t,x_3) \nonumber \\
c_3(x_2) \to c_3(t,x_2). 
\end{eqnarray}
The time dependence of the state is not so clear.
However, we use this simple enlargement because we seem static restriction as only the assumption.
First of all we start with metric as 
\begin{eqnarray}
q_{ab}=\begin{pmatrix}
q_{11}(t,x_2,x_3) & 0 & 0 \\
0 & q_{22}(t,x_1,x_3) & 0 \\
0 & 0 & q_{33}(t,x_1,x_2).
\end{pmatrix}
\end{eqnarray}
Because the formulation is same,
we obtain the state as
\begin{eqnarray}
\cos (2\int c_1(t,x_2,x_3)\delta a_1+\int c_2(t,x_3)\delta a_2 \nonumber \\
+\int c_3(t,x_2)\delta a_3 ) 
\end{eqnarray}
The superposition is also solution as,
\begin{widetext}
\begin{eqnarray}
|\Psi^4(q)\rangle =\cos (2\int c_1^{(1)}(t,x_2,x_3)\delta a_1+2\int c_2^{(1)}(t,x_3)\delta a_2 
+2\int c_3^{(1)}(t,x_2)\delta a_3 )  \nonumber \\
+\cos (2\int c_1^{(2)}(t,x_3)\delta a_1+2\int c_2^{(2)}(t,x_1,x_3)\delta a_2 
+2\int c_3^{(2)}(t,x_1)\delta a_3 )  \\
+\cos (2\int c_1^{(3)}(t,x_2)\delta a_1+2\int c_2^{(3)}(t,x_1)\delta a_2 
+2\int c_3^{(3)}(t,x_1,x_2)\delta a_3 )
\end{eqnarray}
\end{widetext} 
The above state represents the inhomogeneous mini-superspace model with $\phi_{i,i}=0\ \ {\rm for} \ \ i=1,2,3$.

We should discuss classical correspondence.
The averaged value of $q_{ii}$ at each point is given by the integration as
\begin{eqnarray}
\int q_{ii}\cos (2 c_1(t,x_2,x_3) a_1+2 c_2(t,x_3) a_2 \nonumber \\
+2 c_3(t,x_2) a_3 )\prod_i dq_{ii}.
\end{eqnarray}
Then we know $q_{ii}$ is function of $c_i^2(x_j,x_k)$.
So this state has classical correspondence to the mini-superspace Eq.(36).
This spacetime has a classical correspondence if we assume $q_{ii}$ is only depend on $t$.
Then it is corresponding Friedmann universe.
If we treat a Friedman universe mini-superspace model without cosmological constant,
there is no such cosine wave.
Or we treat Bianchi I spacetime without cosmological constant.
The corresponding state is
\begin{eqnarray}
\cos (2\sum_i\int c_i(t)\delta a_i).
\end{eqnarray}
So we can find a cosine wave in the flat without cosmological constant.

If we write the above state $|\Psi ({\rm E.S.S})\rangle $(E.S.S means enlarged static solution),
we can simplify the general Hamiltonian constraint of the LQG.
If furthermore we enlarged this state as,
\begin{eqnarray}
|\Psi ({\rm E.S.S})\rangle \to f[\phi_1,\phi_2,\phi_3]|\Psi ({\rm E.S.S})\rangle ,
\end{eqnarray}
then the Hamiltonian constraint become simple.
Or if we enlarge diffeomorphism constraint as,
\begin{eqnarray}
\sum_{j\not= i}\phi_{i,jj}=c_i(t,x_1,x_2,x_3),
\end{eqnarray}
then the Hamiltonian constraint become simple as
\begin{eqnarray}
\sum_{ij}\frac{1}{2}\frac{\delta^2}{\delta\phi_i\delta\phi_j}+\sum_i (c_i+\oint c_idl \phi_{i,i})e^{\phi_i}=0.
\end{eqnarray}
where $l$ is the loop of $x_j,x_k$.
Then above replaced Hamiltonian constraint is solved.
And this Hamiltonian constraint commute with static restriction.
However, the enlargement of the diffeomorphism constraint is further work.
\section{Conclusion and discussions}\label{sec5}
We can success to solve some inhomogeneous spacetime that is $\phi_{i,i}=0$ mini-superspace.
The functional form i.e. cosine wave is not so important but the ordinary differential equation is important.
The state seems to contain coordinate directly, however by the functional integration these coordinate does not appear in the state.
So the amplitude of the state is independent by each point $t,x_i$.
The state should be a functional of only $g_{ii}$. 

In the local quantum gravity there remain the problem of the boundary condition.
Because of simplification of the Wheeler-DeWitt equation, we only use a diagonal metric components.
However, it produces other difficulty that is the boundary condition problem.

We should comment on the static state more strictly.
First we quantize a inhomogeneous spacetime with static metric and next time we enlarge this state with the support of dynamical metric.
However, the enlarged state is also static state, otherwise the time appears in the state.
So we can say the static diagonal metric space quantize to the state with time evolution.
And some dynamical metric space quantize to static state.
The dynamics of the metric as support and dynamics of the state is different. 

We comment on the diffeomorphism constraint equation.
Because the Hamiltonian constraint is second order derivative,
first order derivative or diffeomorphism constraint may be integration by $q_{ii}$.
Then the diffeomorphism constraint equation become Stokes integration and additional diffeomorphism equation does not appear.
However, in this paper we use an approximation we can obtain the diffeomorphism equation.

Our next work is to solve the Eq.(44) and the discussion of the enlargement of diffeomorphism constraint.
Or we can consider the cosmological constant and then we can consider the C.M.B.
We think the obtained cosine wave is one of the fluctuations of the C.M.B.

\end{document}